\documentclass[english]{article}
\usepackage{float}
\usepackage{babel}
\usepackage{amsmath}
\usepackage{graphicx}
\usepackage{authblk}
\usepackage[letterpaper, portrait, margin=1in]{geometry}

\title{Self-organized novelty detection in driven spin glasses}
\author[1]{Jacob M. Gold \thanks{jacobmg@mit.edu}}
\author[2, 3]{Jeremy L. England}
\affil[1]{Department of Mathematics, Massachusetts Institute of Technology, Cambridge, Massachusetts 02139, USA}
\affil[2]{Physics of Living Systems, Department of Physics, Massachusetts Institute of Technology, 400 Tech Square, Cambridge, MA 02139, USA}
\affil[3]{GlaxoSmithKline AI/ML, 200 Cambridgepark Drive, Cambridge, MA 02140, USA}
\date{}

\begin{document}

\maketitle

\begin{abstract}

We consider a glassy system of interacting spins driven by continual switching amongst a finite set of nonuniform external fields. We find that the system evolves over time towards configurations that minimize the work absorbed from this external drive. The configurations which achieve this are specific to the details of the external fields used to drive the system, and therefore act effectively as a self-organized novelty-detector that embodies accurate predictions about the typical future of its external environment.

\end{abstract}

\section{Introduction}

A variety of many-body systems absorb energy from a given external drive in a way that depends on their configuration.  Both in simulation and experiment, an increasing number of examples have been identified where this variable receptivity to drive work leads to a dissipative adaptation effect: whether in randomly-wired mechanical oscillators \cite{kachman2017self, kedia2019drive}, disordered chemical networks \cite{horowitz2017spontaneous}, or phononic and photonic resonator arrays \cite{bachelard2017emergence, ropp2018dissipative}, the driven steady-state behavior exhibits a self-organized fine-tuning to the particular pattern of the external energy source. 

Spin glasses, which are binary degrees of freedom interacting with each other via quenched, disordered couplings \cite{edwards1975theory, fisher1988nonequilibrium}, are a particularly interesting form of condensed matter to study in this context for two reasons.  Firstly, experimental studies of spin glass materials have demonstrated a flexible memory effect, so that the particular nonequilibrium time-course of cooling can be detected by measuring ac response properties during reheating \cite{jonsson1999nonequilibrium, keim2019memory}. Secondly, analytical models of spin glasses served as the inspiration for the Hopfield net \cite{hopfield1982neural}, which is now the most successful and broadly-applied physical model of learned associative memory \cite{nasrabadi1991object, tatem2001super, park1993economic}.

The classic Hopfield approach involves a training procedure that optimizes the couplings in what is effectively the Hamiltonian of a spin glass model to produce the right relationship between a particular choice of external field (input) and a particular target low-energy spin configuration (output) favored in a low-temperature thermal equilibrium.  The experimentally observed memory effect in spin glasses away from equilibrium, however, points to the possibility that learning behavior can arise in spin glasses through the slow relaxation of glassy coarse-grained features, without any explicit training scheme that modifies the parameters of the Hamiltonian.  The promise of this possibility is strengthened by recent arguments that slow degrees of freedom in driven many-body systems experience pressure to fine-tune themselves in order to reduce work absorption by fast variables \cite{chvykov2018least}.

In this study, we investigate fine-tuned memory effects in simulations of random spin glasses subjected to time-varying external fields.  We demonstrate a tendency of such systems to settle into attractor states that reduce work absorption in a manner specific to the pattern of driving, such that a change in the drive fingerprint leads to a transient rise in dissipation.  Ultimately, we show that this tendency gives rise to an emergent predictive capacity, whereby the current state of the system comes to contain accurate information about the external drive's likely future.

\section{Model}

We start by considering a collection of binary spins $s_i$, each of which has value $\pm 1$, indicating its orientation. These spins interact with one another as well as with an external magnetic field $\mu^{(k)}$ which varies in orientation and magnitude with the spin index. The energy of these spins is given by the classic spin glass Hamiltonian

\begin{equation}
E = \sum_{i, j} J_{ij} s_i s_j + \sum_i \mu_i^{(k)} s_i 
\end{equation}

In order to model thermodynamically-consistent dynamics in such a system we assume that the probability rate per second for any single spin to flip orientation follows an Arrhenius Law of the form 

\begin{equation}
\omega_i = e^{\beta(E_i - B)}
\end{equation}
where $\beta$ is the inverse temperature, $B$ is a constant transition state energy for all flips, and
\begin{equation}
E_i = \sum_j J_{ij} s_i s_j + \mu_i^{(k)} s_i
\end{equation}
This form for the rate ensures local detailed balance and essentially sets a timescale on which spin flips are likely to occur via the choice of $B$. 

Our interest is in a regime of behavior where external driving has a strong influence on spin flip transitions, meaning that far fewer transitions would occur in the absence of a time-varying field.  Practically, this requirement means that the thermal energy scale $1/\beta$ be comparable to or smaller than the energy scale of a typical spin flip. We begin simulating each trajectory at such a temperature from a random initial configuration so that it is quenched in to a nearby local energy minimum  . To introduce a non-equilibrium drive into the model, we consider a set of nonuniform external magnetic fields $\{\mu^{(k)}\}$. Each instance of a field, indexed by a particular value of $k$, is a vector of influences acting on each spin to drive them in the simulation.  As a whole, each field is quite distinct from every other in the same set because the individual elements of the field vector are chosen randomly from a normal distribution, so that every index $k$ corresponds to an external field with a unique fingerprint determined by which spins it is pushing up or down with varying strength. 

By switching which $\mu^{(k)}$ is being applied to the system at fixed time intervals, we provide the opportunity for work to be done on the system by the change in driving field. At each of these fixed intervals the field index which is applied is selected uniformly and randomly from a fixed set of fields whose fingerprints were randomly generated at initialization.
\begin{equation*}
\mu^{(1)} \rightarrow \mu^{(3)} \rightarrow \mu^{(2)} \rightarrow \mu^{(2)} \rightarrow \mu^{(3)} \rightarrow \mu^{(1)} \rightarrow \mu^{(1)} \rightarrow \mu^{(1)} \rightarrow \dots
\end{equation*}
By tracking kinetic and energetic properties of this driven, many-body interacting system, it is possible to search for signatures of adaptation behavior.

\section{Results}

We simulate trajectories of realizations of the model described above with the Gillespie algorithm. The non-zero couplings $J_{ij}$ form an Erdos-Renyi random interaction network \cite{erdHos1960evolution} with mean degree 4, and the value of these couplings are drawn from a standard normal distribution. We consider the low temperature regime $\beta = 3$ where simulating the system for accessible time-scales will typically result in a trajectory where the system relaxes to a local energy minimum at higher energy than the global energy minimum. We choose $B = 4.5$. The site-specific values of each $\mu^{(k)}$ are drawn from a normal distribution with zero mean and $\sigma = 2$.

Each trajectory is produced by starting from a uniformly random arrangement of spins and allowing the system to relax with zero field applied for 100000s. The drive is then turned on and a field $\mu^{(k)}$ is applied to the system. This field changes to another field every 100s.

Observations of least-rattling in other systems have suggested that in a driven system with explicit timescale separation in certain degrees of freedom, the system will prefer configurations that minimize the impact of the fast degrees of freedom on the slow degrees of freedom \cite{chvykov2018least}. In our model there is no explicit timescale separation between degrees of freedom; all spins have to jump the same energetic barrier to flip. For our choice of $\beta$ and $B$, a spin at $E = 0$ would flip once every $~10^6$s. Switching fields every 100s allows for the possibility that some spins may reorient on a much shorter timescale. This allows for the possibility that spins which aren't strongly driven to find arrangements that minimize the rattling from those that are strongly driven. Furthermore, spins that are strongly affected by the drive may reorient in a way that causes them to be less strongly driven, which will in turn reduce rattling. We start by measuring macroscopic quantities that we would expect to change if these effects are present, then investigate the microscopic properties that allow the system to modulate these effects.

\begin{figure}[t]
    
	\centering
	\includegraphics[height=0.5\textwidth]{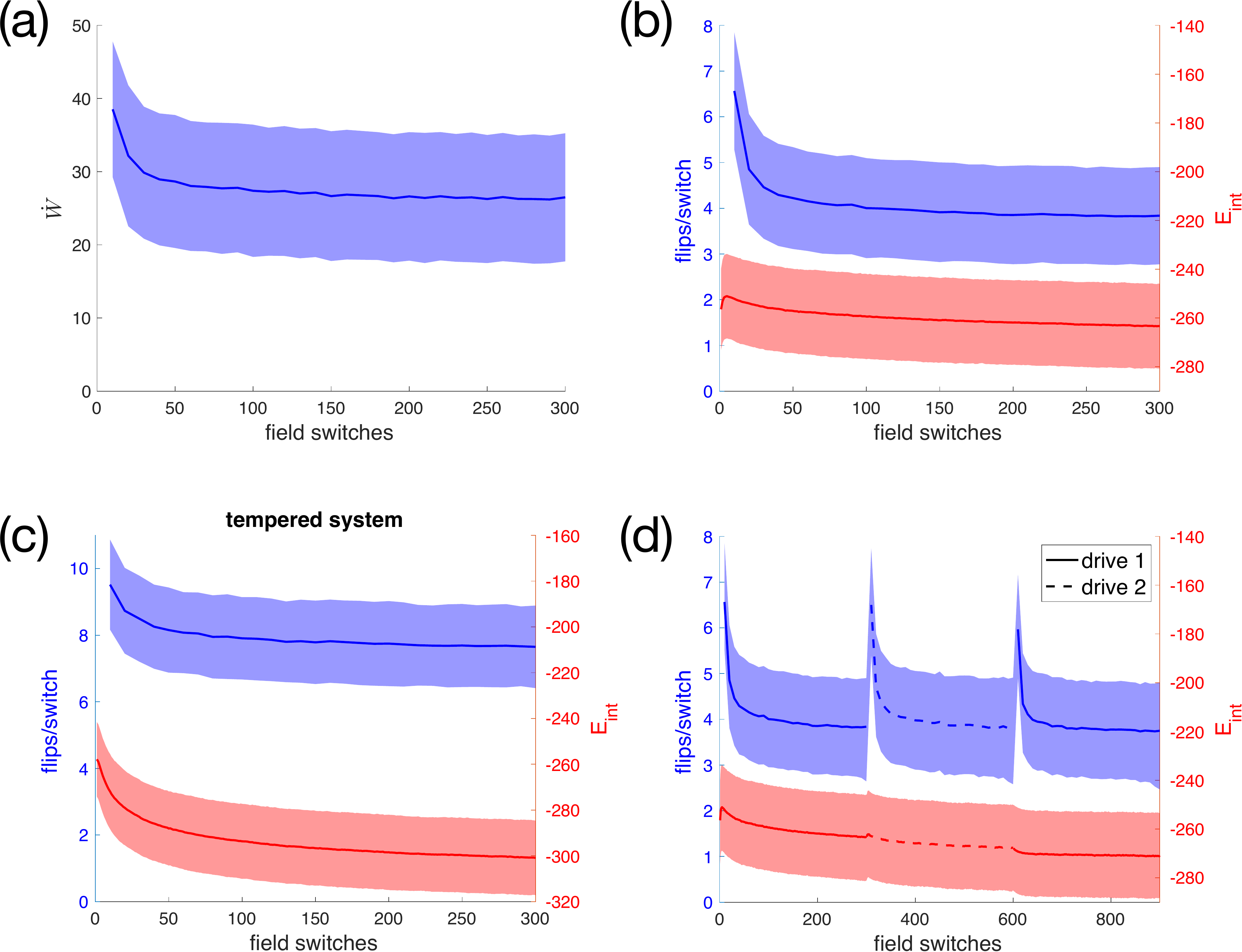}
	\caption{(a) The work absorbed per field switch as measured by the mean of 3000 driven trajectories. The initially high work absorption rate relaxes to a lower value as the system continues to experience the same drive. (b) The initially high flip rate relaxes to a lower value as the system continues to experience the drive. The internal energy of the previously relaxed system is at first increased by the drive but eventually reaches a lower value than the system had relaxed to initially. (c) Increasing the temperature ($\beta = 1.5$) allows the system to further minimize its internal energy, which in turn decreases the flip rate. (d) The mean flip rate and internal energy of the system when a sequence of drives is applied. After 300 field switches the set of fields use to drive the system is changed to another randomly realized set of fields. The flip rate increases back to the value it was at when the first drive was initially applied, before decreasing to the same asymptotic value. Switching back to the first drive after 600 total field switches results in another increase in the flip rate, though lower this time than either of the previous two peaks. The internal energy increases when switching from drive 1 to drive 2, before decreasing to a lower value than was reached during the application of drive 1. Switching back to drive 1 results in a further decrease of internal energy, without any transient increase.}
	\label{fig:driven_work_abs}
	\label{fig:driven_flip_energy}
	\label{fig:tempered_flip_energy}
	\label{fig:change_drive_flip_energy}
\end{figure}

The work absorption rate is a measure of how strongly the system is driven by the external fields. The work absorption is defined as the change in energy of the system any time the external fields are changed
\begin{equation} \label{eq:work_abs}
\Delta W = \sum_{i} (\mu_i^{k'} - \mu_i^{k}) s_i
\end{equation}
To calculate the work absorption rate we measure the change in total work absorbed in a certain window of time and divide by the duration of that window. The work absorption rate declines over time before plateauing at a value lower than its initial rate (Fig. \ref{fig:driven_work_abs}(a)).

Next we turn to investigating the mechanism that determines the rate of work absorption of a particular configuration. While it is true that the measured work absorbed (Eq. \ref{eq:work_abs}) changes instantaneously at the moment the external magnetic field changes, spins must flip in between shifts in the field in order for a net positive amount of work to be absorbed on average. This suggests a strong correlation between the rate at which work is absorbed and the rate at which spins in the system flip. At each moment, the energy of each spin sets the rate at which that spin will flip. One might then expect that the internal energy of a configuration
\begin{equation}
E_{\mathrm{int}} = \sum_{i,j} J_{ij} s_i s_j
\end{equation}
would be an important component in the rate at which spins flip, and therefore a good predictor of the rate at which spins flip. We find that both the spin flip rate and the internal energy of the driven system approach lower values in the long-term than they were at initially (Fig. \ref{fig:driven_flip_energy}(b)).

This behavior in which giving the system access to a larger amount of energy influx allows it to reach a lower internal energy is similar to a tempering process where jumping to a new higher fixed temperature allows it to do the same thing \cite{marinari1992simulated}. To make this comparison we construct a system in the same way except instead of turning on the drive at $t = 0$, we change the inverse temperature from $\beta = 3$ to $\beta = 1.5$. The result is that the spin flip rate and internal energy of the system both decrease (Fig. \ref{fig:tempered_flip_energy}(c)). At a glance the qualitative behavior of the driven system and the tempered system appears to be the same. However, the relative decrease in internal energy of the tempered system is much more dramatic than in the driven system, and yet the reduction in flip rate is smaller. (Note that there is not a measurable quantity in the tempered system analogous to the work absorption rate in the driven system.) This suggests there is another mechanism responsible for the majority of the reduction in spin flip rate and work absorption rate in the driven system.

Because there is a clear difference in the response of the system to driving and tempering, we also ask how specific is the response is to the details of the particular realization of the drive we apply. After driving the system by switching 300 times between fields within a particular set of fields, we draw a new set of fields from the same distribution and switch between those to drive the system instead. This change results in an increase of the spin flip rate to the same value it was at when the first drive was initially applied (Fig. \ref{fig:change_drive_flip_energy}(d)). This suggests that the types of configurations that minimize work absorption are specific to the particular realization of the drive. Changing the drive back to initial set of fields results in a smaller increase in the flip rate, indicating that the system retains some memory of the first drive after the second has been applied.

    

\begin{figure}[t]
    
	\centering
	\includegraphics[height=0.5\textwidth]{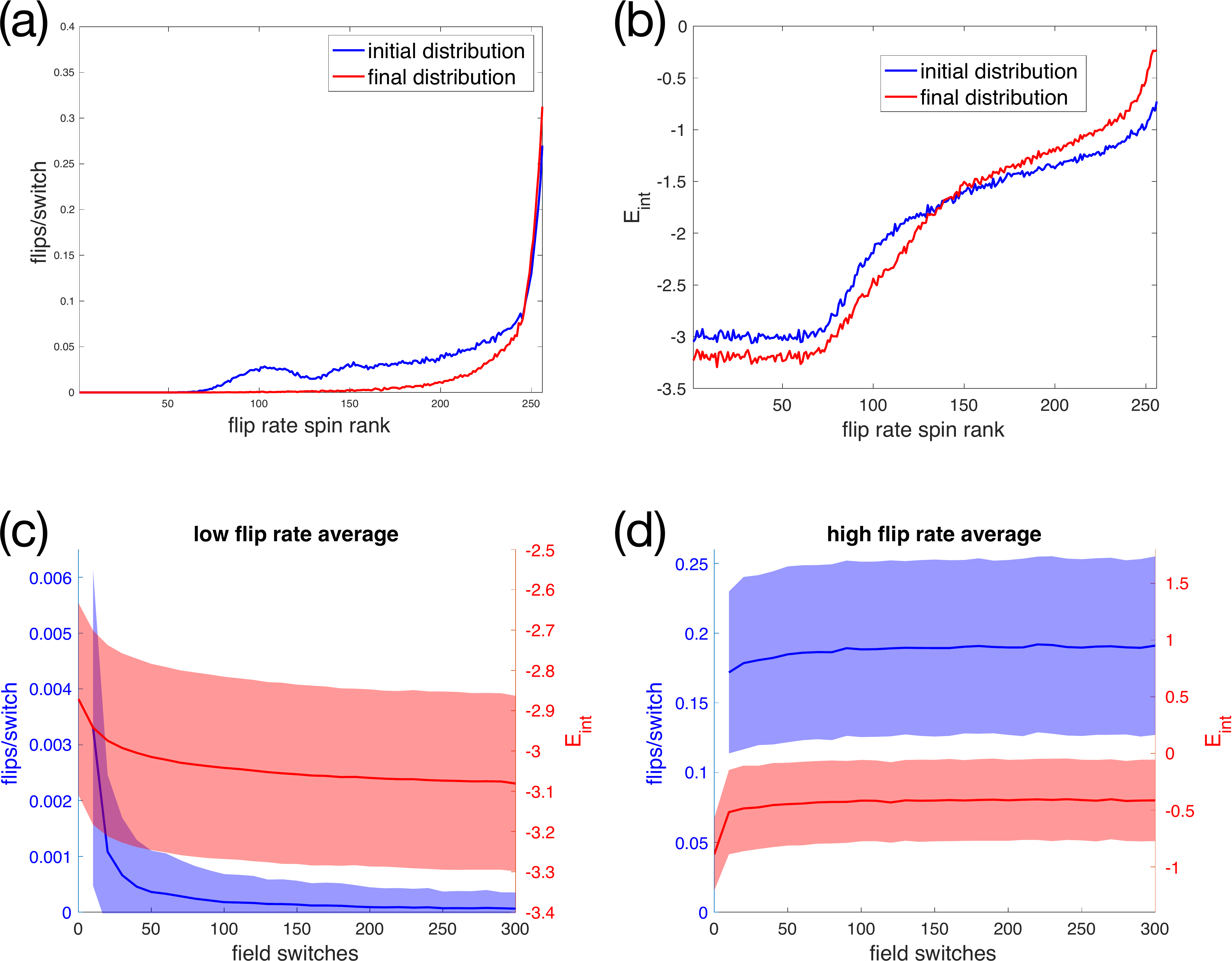}
	\caption{(a) The mean flip rate of spins in the first and final five field switches of a trajectory, where before averaging spins are first sorted from low to high by the total number of flips across that trajectory. The majority of the total number of flips is contributed by just a few spins. The majority of the reduction in spin flip rate is accounted for by those spins which flip an intermediate number of times. (b) The mean internal energy of individual spins, where before averaging spins are first sorted by the total number of flips across that trajectory. Those spins that flip the least tend to decrease in energy, while those that flip the most increase in energy. (c) The average flip rate and internal energy of the 10 spins which over the course of a trajectory flip the most. Contrary to the behavior of the average, these spins experience an increase over time in both their internal energy and flip rate. (d) The average flip rate and internal energy of the 100 spins which over the course of a trajectory flip the least.}
	\label{fig:spin_rank_flip_counts}
	\label{fig:spin_rank_int_energy}
	\label{fig:driven_high_flip}
	\label{fig:driven_low_flip}
\end{figure}

Since the internal energy doesn't explain entirely the work absorption rate and spin flip rate of the driven system, we consider what the possible mechanisms are that allow the system to modulate those quantities.

Given the number of components of the system that are randomly realized, there isn't any reason to expect that every degree of freedom in the system contributes in the same way to these quantities measured in the system as a whole. We consider these same quantities measured for individual spins, where the spins are sorted based on the total number of times they flip over the course of a trajectory before the quantities are averaged across realizations. We find that the majority of the spin flips in the system as a whole are performed by the few spins which flip the most (Fig. \ref{fig:spin_rank_flip_counts}(a)). Following the time trace of just these spins reveals that contrary to the mean behavior of the system, their flip rate increases over time, as does their internal energy (Fig. \ref{fig:driven_high_flip}(c)). There are many more spins though that experience a drastic reduction in flip rate, as well as a reduction in internal energy (Fig. \ref{fig:driven_low_flip}(d)).

While spin flips in the driven system are also mediated by stochasticity from the thermal bath, the primary driver of flipping is the work absorbed from the changing external field. Decreasing the interaction energy of a driven spin will reduce the likelihood that a given field will be strong enough to cause it to flip, but there are other mechanisms that can change the rate at which a spin flips. If all of the fields at a specific site favor a spin pointing in a certain direction, one of two things will happen. The neighbors of this spin might also rearrange so that the direction favored by the external fields also minimizes the interaction energies; once the spin is arranged favorably with both the external fields and its neighbors it is unlikely to flip again, resulting in a decrease in its flip rate. Alternatively the external fields might stabilize an otherwise less favorable higher energy state. In this case it is still possible for the spin to flip, but it does so at a slower rate over time.

\begin{figure}[t]
    
	\centering
	\includegraphics[height=0.25\textwidth]{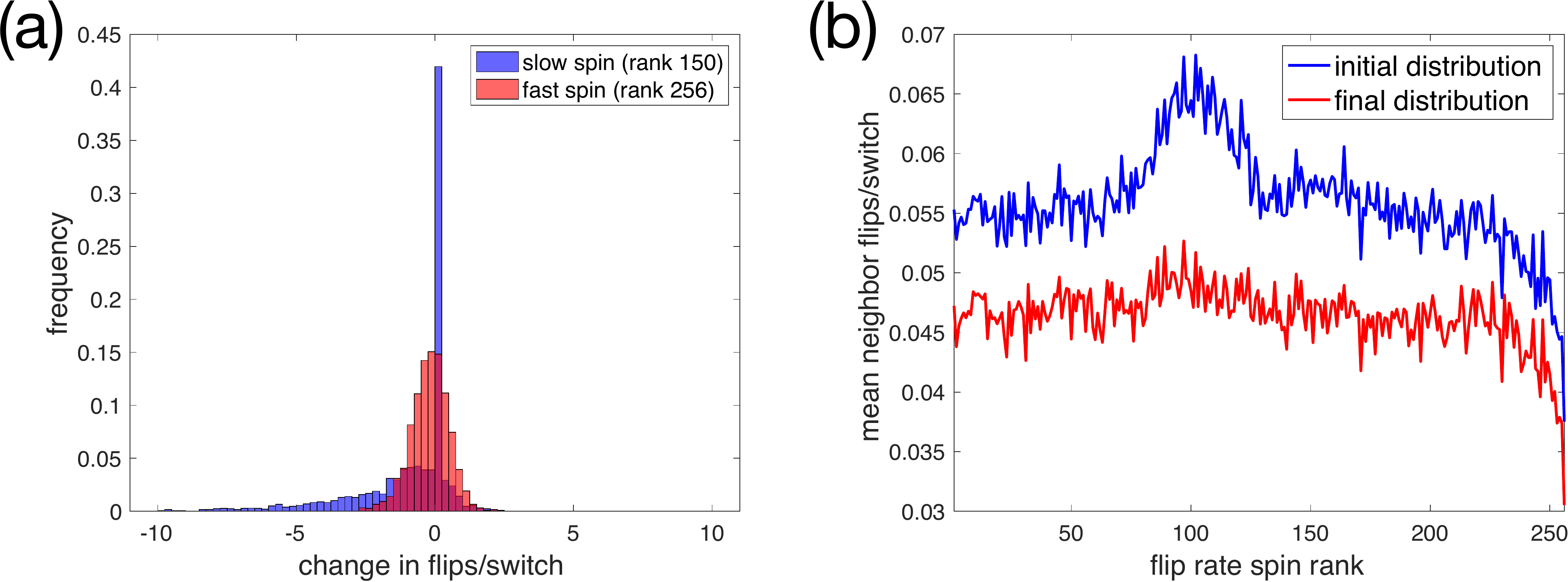}
	\caption{(a) The difference in total flip rate of the rest of the system when a particular spin is oriented up or down. The difference is calculated by subtracting the flip rate of the more likely orientation from the flip rate of the less likely orientation; a negative mean difference indicates that spins spend more time in arrangements where the rest of the system flips less. A fast spin which flips a lot over the course of a trajectory is less likely to have a large impact on the overall flip rate of the system than a slow spin which only flips occasionally. (b) The mean flip rate of the neighbors of each spin ranked by how many times that spin flips over the course of a trajectory, with rank 256 being the spin that flips the most. The spins which flip the most have neighbors which flip the least, while the spins whose neighbors flip the most tend to flip about once over the course of a trajectory.}
	\label{fig:spin_significance}
	\label{fig:neighbor_flip_rate}
\end{figure}

These spins that increase in energy (Fig. \ref{fig:spin_rank_int_energy}(b)) that only flip occasionally (Fig. \ref{fig:spin_rank_flip_counts}(a)) can be thought of as slow dynamic degrees of freedom. The increase in energy and decrease in flipping rate is explained by the alignment of multiple strong external fields ([moved to supplement]]). The drive is usually able to keep the spin pinned in a high energy state, but when they do flip, they exert a larger influence on the total flip rate of the configuration than the spins which flip the most (Fig. \ref{fig:spin_significance}(a)). Additionally they spend more time in states where the total flip rate of the system is lower.

Sometimes spins can be observed to increase the rate at which they flip. The interaction energy of these spins tends to increase over time, making it easier for the spin to flip in the presence of a field of fixed strength. The neighbors of these spins also flip less than the average spin (Fig. \ref{fig:neighbor_flip_rate}(b)); their stability allows the fast flipping spins to flip without transmitting energy that causes the remainder of the system to rearrange. Spins that flip approximately once over the course of a trajectory have neighbors which flip the most on average; either because they are stabilized by a strong external field or driven to reach lower energy with respect to their neighbors they become more stable, allowing their neighbors to flip freely without disturbing the rest of the system configuration.

\section{Discussion}

In the above, a random spin glass is driven by a time-varying external force that is constrained to a very limited sub-space of all possible values of external forcing on each spin.  In essence, this means that the external drive has a particular pattern or fingerprint to it, which it might in principle possible to learn or recognize.  The above results demonstrate that, simply by undergoing normal physical dynamics in the presence of this patterned input, the initially na\"{i}ve random spin glass does learn to distinguish between drive fingerprints, so that its work absorption rate serves as a novelty-detector that recognizes the difference between a new drive and a familiar one.  Moreover, by analyzing the flipping rates of different spins in the network, it is possible to show that this self-organized fine-tuning originates in the kinetic slowdown and stabilization of a subset of spins that exert an outsize influence on the flipping rates of all the others. 

Even though our model does not contain explicit timescale separation between different spins, in practice spins flip at a wide range of rates (Fig. \ref{fig:spin_rank_flip_counts}(a)), providing an effective timescale separation. The slowest spins are those which have their energy reduced the most, insulating them from changes in their neighbors' orientations due to the drive, reducing rattling.

With this effective timescale separation, slow spins have a larger impact on the total flip rate of the system than fast spins. Fast spins are insulated from their neighbors, so changes in their orientation have a smaller impact on their neighbor's flipping rate (Fig. \ref{fig:spin_significance}(a). Slow spins flip more rarely, so there is less reason to favor configurations that minimize the impact of such a flip, resulting in larger changes in flip rate when they do occur. One can consider these slow degrees of freedom to be controllers for the fast degrees of freedom, which self-organize to arrangements that minimize the rattling {chvykov2018least} that occurs in the system.

Which degrees of freedom are fast and slow are determined by the specific realizations of the interactions and driving fields. Spins with fewer neighbors are more likely to be fast spins, since they have higher energy and their flipping has a smaller impact on the rest of the system. The magnitude of the external fields still has to be above a certain threshold to drive a spin enough that it will flip at a high rate. Those slow spins that still flip occasionally and exert a large influence on the total flip rate of the system are dependent on having multiple aligned external fields that allow them to remain at high energy. Because the effective timescales are highly dependent on the details of the drive, the configurations that minimize the work absorption rate are also finely tuned to the details of the drive (Fig. \ref{fig:change_drive_flip_energy}(d)).

\section{Conclusion}

We have demonstrated that a spin glass driven by time-dependent external fields will evolve to configurations that absorb less work from that external drive. There are a variety of mechanisms through which spins can modulate their flipping rate to minimize the impact of fast spins on the slower spins in the system. These mechanisms are specific to the details of the external fields used to drive the system; changing these fields results in a resetting of the system as if it had just relaxed to its current state.

While the basis of the predictions we made about this model were inspired by a framework with explicit timescale separation between different degrees of freedom, our results show the same type of behavior is possible in a system where in the absence of a fast external drive, all degrees of freedom are equally slow, but once the drive is applied they can self-organize in glassy fashion and operate on a range of timescales. Because the model we chose to investigate is generic beyond having many driven degrees of freedom moving through a rugged energy landscape in configuration space, we expect these results to generalize to many such systems. Future analysis of experimental systems with these properties would serve to investigate that possibility. Because the evolution of these systems is specific to the details of the drive, it may be possible to select for states in that system by picking a drive where work absorption is minimized in that state.

\section*{Acknowledgements}

JMG is funded by the Air Force Office of Scientific Research under Grant FA9950-17-1-0136. This material is based upon work supported by, or in part by, AFOSR Grant FA9550-19-1-0411. JLE has been funded by the AFOSR Grant FA9950-17-1 and by the James S. McDonnell Foundation Scholar Grant 220020476.

\bibliographystyle{plain}
\bibliography{refs}

\end{document}